\documentclass[reprint, aps,showpacs, prl, floatfix, superscriptaddress]{revtex4-1}

\usepackage{graphicx}
\usepackage{dcolumn}
\usepackage{bm}
\usepackage{hyperref}
\usepackage{epstopdf}
\usepackage{CJK}
\usepackage{xcolor}

\begin{document}
\begin{CJK*}{GBK}{} 

\preprint{APS/123-QED}

\title{Continuously tunable electronic structure of transition metal dichalcogenides superlattices
}

\author{Yong-Hong Zhao}
 \email{yhzhao@sicnu.edu.cn}
 \affiliation{College of Physics and Electronic Engineering,
              Institute of Solid State Physics,
              Sichuan Normal University, Chengdu 610068, China}
\affiliation{Department of Physics and the Center of Theoretical and Computational Physics,
The University of Hong Kong, Hong Kong, China}

\author{Feng Yang}
\affiliation{College of Physics and Electronic Engineering,
              Institute of Solid State Physics,
              Sichuan Normal University, Chengdu 610068, China}
\affiliation{Department of Physics, Renmin University of China,
              Beijing 100872, China}%

\author{Jian Wang}
\affiliation{Department of Physics and the Center of Theoretical and Computational Physics,
The University of Hong Kong, Hong Kong, China}

\author{Hong Guo}
\affiliation{Centre for the Physics of Materials and Department of Physics, McGill University, 3600 rue University, Montreal PQ, Canada H3A 2T8}

\author{Wei Ji}
 \email{wji@ruc.edu.cn}
 \affiliation{Department of Physics, Renmin University of China,
              Beijing 100872, China}%

\date{\today}

\begin{abstract}
Two dimensional transition metal dichalcogenides (TMDC) have very interesting properties for optoelectronic devices. In this work we theoretically investigate and predict that superlattices comprised of MoS$_2$ and WSe$_2$ multilayers possess continuously tunable electronic structure having direct band gap. The tunability is controlled by the thickness ratio of MoS$_{2}$ versus WSe$_{2}$ of the superlattice. When this ratio goes from 1:2 to 5:1, the dominant K-K direct band gap is continuously tuned from 0.14 eV to 0.5 eV. The gap stays direct against -0.6\% to 2\% in-layer strain and up to -4.3\% normal-layer compressive strain. The valance and conduction bands are spatially separated. These robust properties suggest that MoS$_2$ and WSe$_2$ multilayer superlattice should be an exciting emerging material for infrared optoelectronics.
\end{abstract}

\maketitle
\end{CJK*}

It has been shown recently that monolayer (ML) transition metal dichalcogenides (TMDC)
have very interesting properties as emerging materials for optoelectronic devices \cite{vdwHetero,naturenano2011,naturenano2012,chemRev,RPP2013,strainpku,hetero2013}.
These ML-TMDCs have direct band gaps in the visible frequency range and strong spin-orbit coupling (SOC). Prototypes of photoactive devices made by these materials already offer quantum efficiency of 30\% \cite{heteroLight} and theoretical power conversion efficiency up to 1\% in solar cell applications\cite{TMDCpv}. Interestingly, several TMDCs are indirect gap material in the bulk form but are direct gap material as ML. These materials include MoS$_2$, MoSe$_2$, WS$_2$, and WSe$_2$ \cite{mos2mono,wse2mono}.

To realize a powerful TMDC optoelectronics, a most interesting scenario is to somehow make the band gap tunable. While strain inside the layers of TMDC can linearly decrease the band gap of monolayer MoS$_2$ \cite{MoS2strain,smt}, it also makes the band gaps of ML-TMDC indirect\cite{strainmos2}. The band gaps of hetero-bilayers, for instance a ML WS$_2$ stacked on ML WSe$_2$, are predicted to be direct\cite{hetero}. But when the hetero-bilayers periodically repeat to form a superlattice, an {\it indirect} gap again emerges.  It appears that ML-TMDC is paramount for retaining the direct gap\cite{hetero,mos2mono,MoS2strain,smt,strainmos2,wse2mono,mos21995,
mos22009,wse2bulk,wse1997,MoS2_arpes}. This restriction limits the fundamental building blocks for TMDC optoelectronics. A very important open problem is how to design TMDC systems having both tunable electronic structures and direct band gaps. It is the purpose of this work to provide a solution to this problem.

\begin{figure}
\includegraphics[width=8.6cm]{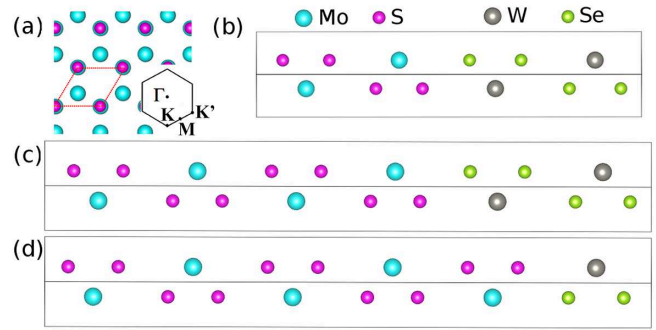}
\caption{\label{fig:struct}(Color online) Crystal structure of representative superlattices, (a) the 1:1 superlattice where MoS$_2$:WSe$_2$ equals to 1:1; (b) the 2:1 superlattice where MoS$_2$:WSe$_2$ equals to 2:1; (c) the 5:1 monolayer superlattice where all MoS$_2$ and WSe$_2$ bilayers were replaced by the corresponding monolayers.}
\end{figure}

\begin{figure}
\includegraphics[width=8cm]{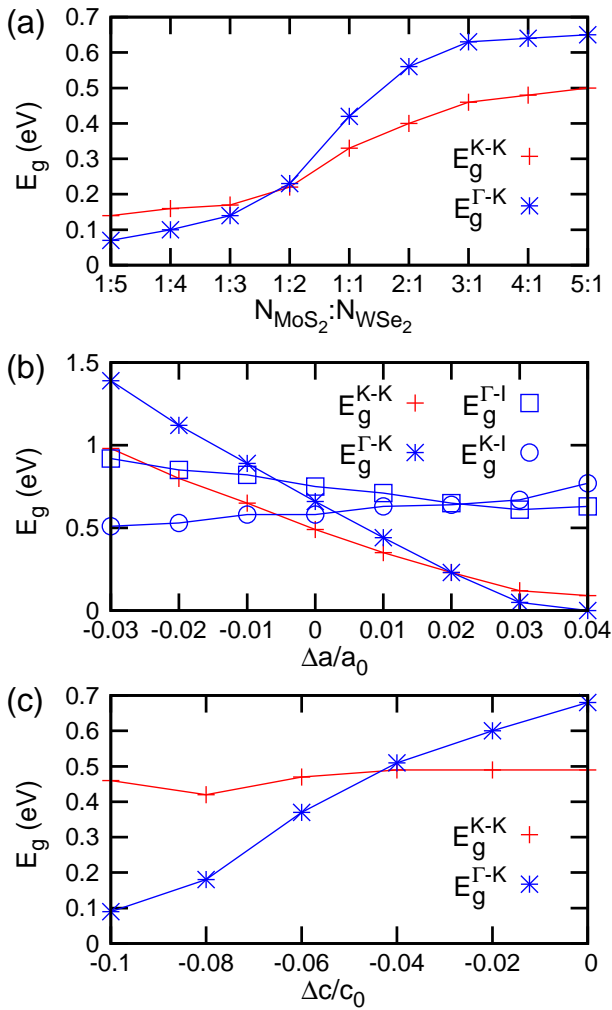}
\caption{\label{fig:gap}
(Color online) Direct (K-K) and indirect ($\Gamma$-K) band gaps of MoS$_2$/WSe$_2$ superlattice as a function of: (a) the MoS$_2$ and WSe$_2$ ratio; (b) the in-layer strain and (c) the normal-layer compressive strain. The red color and the plus symbols represent K-K direct gap; the blue color and the star, square and circle symbols represent the $\Gamma$-K, $\Gamma$-I and K-I indirect gaps, respectively.}
\end{figure}

\begin{table}
 \caption{\label{table:lattice}
 Fully relaxed lattice parameters for MoS$_2$/WSe$_2$ superlattices. N$_{{\textrm {MoS}}_2}$ and N$_{{\textrm {WSe}}_2}$ represent the number of bilayers in each superlattice for MoS$_2$ and WSe$_2$, respectively.}
 \begin{ruledtabular}
 \begin{tabular}{ccccc}
 &N$_{{\textrm {MoS}}_2}$:N$_{{\textrm {WSe}}_2}$ &{a/c (\AA)} &N$_{{\textrm {MoS}}_2}$:N$_{{\textrm {WSe}}_2}$ &{a/c (\AA)}\\
 \hline
 &1:1 &3.248/25.422 &1:1 &3.248/25.422 \\
 &1:2 &3.271/38.478 &2:1 &3.229/37.919\\
 &1:3 &3.281/51.483 &3:1 &3.220/50.326\\
 &1:4 &3.290/64.655 &4:1 &3.219/62.477\\
 &1:5 &3.294/77.600 &5:1 &3.214/74.932
 \end{tabular}
 \end{ruledtabular}
\end{table}

In particular, we report a discovery that extends the candidates of building blocks for TMDC optoelectronic devices from only ML to potentially limitless possibilities, which may open an interesting area of TMDC. By first principles electronic structure theory we show that TMDC {\it multilayer} superlattices offer direct gap and tunable electronic structures. The tunability is achieved by the thickness ratio of the multilayers across the hetero-interface. Namely, the 1:1 MoS$_2$(bilayer)/WSe$_2$(bilayer) superlattice, as shown in Fig. \ref{fig:struct}(b), has a direct band gap of 0.33 eV which is continuously tunable by changing the MoS$_2$ and WSe$_2$ ratio. The direct band gap is also robust against in-layer ($a$ direction) strain from -0.6\% to 2\%, and normal-layer ($c$ direction) uniaxial strain up to -4.3\%. Moreover, the valance and conduction bands were found spatially separated. These interesting results strongly suggest that TMDC {\it multilayer} superlattices may well be the emerging optoelectronic device material in the mid-infrared frequency range.

We carry out density functional theory (DFT) calculations using the generalized gradient approximation (GGA) for the exchange-correlation potential\cite{pbe}, the projector augmented wave method\cite{paw,paw2}, and a plane wave basis set as implemented in the Vienna {\it ab-initio} simulation package\cite{vasp,vasp2}. The SOC correction, calculated to be about 160 meV and 400 meV for MoS$_2$\cite{mos2mono,monosoi} and WSe$_2$\cite{wse2soi,monosoi} respectively, was included in our analysis which strongly influences the band structure of material. The energy cutoff for plane-wave basis was set to 400 eV for all calculations. A $k$-mesh of 24 $\times$ 24 $\times$ 1 or 24 $\times$ 24 $\times$ 3, depending on different values of lattice parameter $c$, was adopted to sample the first Brillouin zone (BZ). In geometry optimization, dispersion forces were considered by employing the semi-empirical PBE-D2 method\cite{vdw1} which was found highly suitable for MoS$_2$\cite{vdw2}. The shape and volume of each supercell were fully optimized and all atoms in it were allowed to relaxed until the residual force per atom was less than 0.01 eV/\AA.

Besides the 1:1 superlattice, two additional families of superlattices were investigated. In the first, the thickness of WSe$_2$ is kept constant, i.e. a bilayer, and the thickness of MoS$_2$ is varied from bilayer to four layers (see Fig. \ref{fig:struct} (c)) and eventually to ten layers. In other words, we increase the MoS$_2$/WSe$_2$ ratio (N$_{{\textrm {MoS}}_2}$:N$_{{\textrm {WSe}}_2}$) from 1:1 to 5:1. In the second family of superlattice, the thickness of MoS$_2$ is kept fixed at bilayer and the ratio decreases from 1:1 to 1:5. Atomic structures including volume, c/a and internal atomic coordinates, of all superlattices were fully optimized and results presented in Table. \ref{table:lattice}. The associated in-layer lattice parameter $a$ varies from 3.214\AA~to 3.294\AA, close to the values of bulk MoS$_2$ and WSe$_2$ respectively, when the N$_{{\textrm {MoS}}_2}$:N$_{{\textrm {WSe}}_2}$ ratio decreases from 5:1 to 1:5.

\begin{figure}
\includegraphics[width=8.6cm]{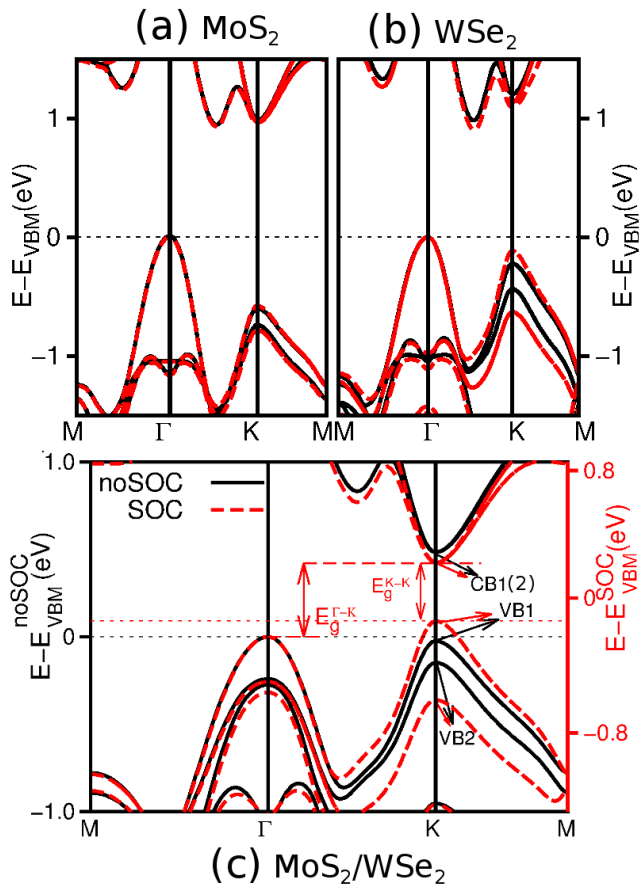}
\caption{\label{fig:band}(Color online) Band structures of (a) bulk MoS$_2$ and (b) WSe$_2$, and (c) the 1:1 superlattice shown in Fig. \ref{fig:struct}(a). The black solid lines are results without SOC; the red dashed lines are results with SOC. The horizontal thin dotted lines mark the energies of VBMs for non-SOC (black) and SOC (red) results. E$^{K-K}_{g}$ and E$^{\Gamma-K}_{g}$ represent direct and indirect gaps of the superlattice, respectively. States CB1, CB2, VB1, and VB2 are the four states around the band gap.}
\end{figure}

\begin{figure*}
\includegraphics[width=16cm]{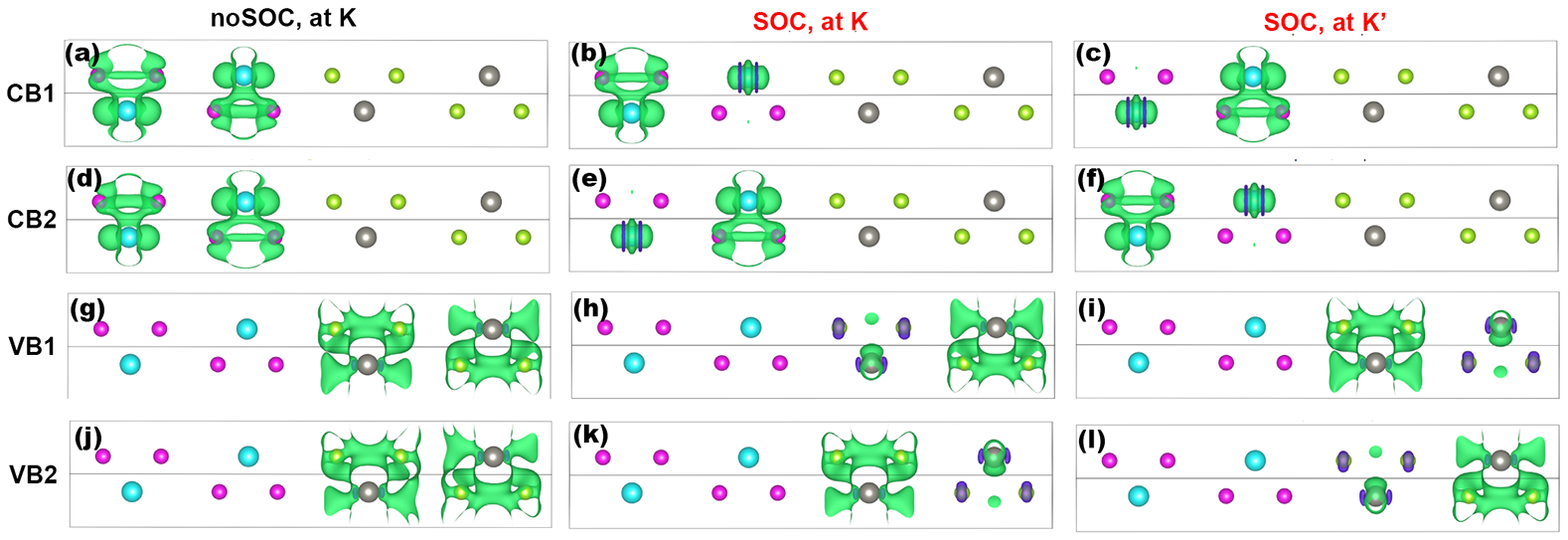}
\caption{\label{fig:pcharge}(Color online) Visualized wavefunction of the 1:1 superlattice, for CB1 (a)-(c), CB2 (d)-(f), VB1 (g)-(i) and VB2 (j)-(l), respectively. Non-SOC result at K and SOC results at K and K' are shown at the left, middle and right panels, respectively. The isosurface is 0.001.}
\end{figure*}

Electronic properties of layered materials usually have a strong thickness dependence\cite{thicknesswse2}. We firstly investigate the evolution of band gaps as a function of the relative thickness of MoS$_2$ and WSe$_2$, i.e. the MoS$_2$/WSe$_2$ ratio.
From the relaxed atomic structures of the superlattice, the band structures were calculated by including the SOC correction. Figure \ref{fig:gap}(a) shows the values of the K-K direct and $\Gamma$-K indirect band gaps as a function of the MoS$_2$/WSe$_2$ ratio.
Both the direct and indirect gaps becomes larger as a function of this ratio. The largest direct (indirect) gap at N$_{{\textrm {MoS}}_2}$:N$_{{\textrm {WSe}}_2}$=5:1 is 0.50 eV (0.65 eV); while the smallest direct (indirect) gap reaches 0.14 eV (0.07 eV) when N$_{{\textrm {MoS}}_2}$:N$_{{\textrm {WSe}}_2}$=1:5. We conclude that both the direct and indirect band gaps are continuously tunable, going from nearly zero-gap to narrow gap semiconductors according to the MoS$_2$/WSe$_2$ ratio of the superlattice. If all bilayers in the 5:1 superlattice are replaced with monolayers, as shown in Fig. \ref{fig:struct}(d), the superlattrice retains the direct band gap of about 0.52 eV. The tuning of the band gaps can be explained by the stacking and SOC induced VB separations. Very briefly, the thinner the WSe$_2$ in the superlattice, the smaller the VB separation at the K point, hence the larger the K-K direct band gap. In the monolayer case shown in Fig. \ref{fig:struct}(d), the stacking induced VB separation is further suppressed, giving rise to an even larger K-K gap. These behaviors will be elucidated further below, and we refer interested readers to Fig. S2(a) of the Supplemental Material\cite{epaps} for more details.

Strain significantly affects band gaps of TMDC\cite{strainmos2,strainpku,strainpku1,straingap} and usually changes the dominant band gap from direct to indirect as discussed above. But our superlattice resists such an undesirable change. As an example, we investigate the strain effect using the 5:1 superlattice which has the largest direct and indirect gaps among all superlattices we studied. Figures \ref{fig:gap}(b) and (c) plot the evolution of the band gap as a function of in-layer and normal-layer strains, respectively. In-layer strain was applied by varying the in-layer lattice constant $a$. Figure \ref{fig:gap}(b) plots four gaps, the K-K direct gap and the $\Gamma$-K, $\Gamma$-I, and K-I indirect gaps, versus $\Delta a$ from -3\% to 4\%. We found that the superlattices are direct gap material dominated by the K-K direct gap (red line) in the range of $-0.6\% \leq\Delta a\leq 2.0\%$. Outside this range, the K-I or $\Gamma$-K indirect gaps are the smallest gaps. Normal-layer compressive strain was applied by the same scheme where the normal-layer lattice parameter $c$ was shortened up to 10\% at a step of 2\%. Values of K-K direct and $\Gamma$-K indirect band gaps were plotted in Fig. \ref{fig:gap}(c). The direct gap is almost a constant but the indirect gap rapidly decreases when the strain increases up to 10\%, resulting to a transition from direct to indirect at a compressive ratio of $\sim$ -4.3\%. This strain corresponds to an external pressure of $\sim$ 3 GPa. We conclude that the proposed superlattices are quite robust against strain as far as the direct gap is concerned.

Having established the properties of tunable electronic structure and direct gaps of the superlattice, we now discuss the origin of the direct band gap by examining the bulk MoS$_2$, WSe$_2$ and the 1:1 superlattice, as shown in Fig. \ref{fig:band}. With or without SOC, the bulk materials have indirect band gaps between $\Gamma$ and a certain intermediate point (I) along $\Gamma$-K, as shown in Fig. \ref{fig:band}(a) and (b), consistent with previous reports\cite{mos21995,mos22009,wse2bulk,wse1997}. The most remarkable change of the bulk band structure induced by SOC is the significant VB splitting at the K point of WSe$_2$ (see Fig. \ref{fig:band}(b)), though this change is minor for MoS$_2$ (Fig. \ref{fig:band}(a)). The non-SOC band structure of the 1:1 superlattice (black solid line in Fig. \ref{fig:band}(c)) appears to be a combination of the bulk band structures of MoS$_2$ and WSe$_2$. Wave function visualization (see Figs. \ref{fig:pcharge} and S1) shows that the two highest VBs (VB1 and VB2) mainly originate from the VBs of WSe$_2$; and the two lowest CBs (CB1 and CB2) are solely contributed by the CBs of MoS$_2$. This combination significantly reduces the K-K band gap, although it is still slightly larger than the $\Gamma$-K indirect gap. The VBM, a mixture of MoS$_2$ and WSe$_2$ states, was found at $\Gamma$ which is 18 meV higher in energy than VB1 at the K. We align the SOC corrected band structure (red dashed line in Fig. \ref{fig:band}(c)) to the non-SOC one by VB1 at the $\Gamma$ point in order to better understand the role of SOC. The SOC correction, similar to the WSe$_{2}$ case, further enlarges the gap between VB1 and VB2 and pushes the slightly split CB1 and CB2 down at the K point, which results in the corrected VBM at the K point rather than at $\Gamma$. Together with the CBM at K, we arrive at a crossover from the indirect $\Gamma$-K gap ($E_{g}^{\Gamma-\mathrm{K}}$) of 0.48 eV to the direct K-K gap ($E_{g}^{\mathrm{K-K}}$) of 0.33 eV.

The microscopic physics is further revealed by plotting the wave functions of CB1, CB2, VB1, and VB2 at the K point. We denote them as $\Psi^{\mathrm{CB1}}_{\mathrm{K}}$, $\Psi^{\mathrm{CB2}}_{\mathrm{K}}$, $\Psi^{\mathrm{VB1}}_{\mathrm{K}}$, $\Psi^{\mathrm{VB2}}_{\mathrm{K}}$, and visualize them by plotting their square norm. Figures \ref{fig:pcharge}(a)-(f) clearly show that $\Psi^{\mathrm{CB1(2)}}_{\mathrm{K}}$ is mainly composed of Mo $d_{z^2}$ and S 3-fold $p$ orbitals. Although $\Psi^{\mathrm{VB1}}_{\mathrm{K}}$ and $\Psi^{\mathrm{VB2}}_{\mathrm{K}}$ are separated in energy, Figs. \ref{fig:pcharge}(g)-(l) explicitly show that they originate from the same combination of in-layer 3-fold W $d$ and Se $p$ states. The results without SOC thus indicate that stacking the materials breaks the energetic degeneracy of VB1 and VB2 but keeps the composition of their wave functions. By including SOC, on the other hand, the composition of the wave functions are also broken as shown by panels $(b,c, e,f,h,i,k,l)$ of Fig. \ref{fig:pcharge}.
The wave function mainly resides at one side of a bilayer at a certain K point, for example, most portion of $\Psi^{\mathrm{VB1}}_{\mathrm{K}}$ locates at the right side of the MoS$_{2}$/WSe$_{2}$ interface, while $\Psi^{\mathrm{VB1}}_{\mathrm{K'}}$ at the left side [Fig. \ref{fig:pcharge}(h,i)]. In other words, SOC makes the originally degenerate K and K' points distinguishable. The enlarged separation moves the VBM from the $\Gamma$ to the K point, bringing about the direct K-K gap in these superlattices.  We conclude that SOC plays the critical role to produce the direct gaps in the proposed superlattices. Finally, we point out that the spatially separated valance and conduction bands is a desirable property since it should help to reduce recombination of the electron-hole pairs and lead to higher quantum efficiency of photoelectron generation.

In summary, we propose a novel MoS$_2$/WSe$_2$ superlattice that offer continuously tunable electronic structures and direct band gaps which are robust against reasonable ranges of external strains. We identify that the robust direct band gap is resulted from a strong spin-orbit coupling in WSe$_2$ and the band alignment between MoS$_2$ and WSe$_2$. Either the MoS$_2$/WSe$_2$ thickness ratio or an additional in-layer external stress can continuously change the direct gap from 0.14 eV to 0.50 eV. The spatially separated valance (on WSe$_2$) and conduction (on MoS$_2$) bands is another very attractive property of the proposed superlattice from the application point of view. On the fundamental side, if valance electrons are excited to the conduction bands by polarized light, the spatially separated VB and CB shall give rise to a magnetic moment normal to the MoS$_2$/WSe$_2$ interface. Finally, the distinguishable K and K' points caused by SOC make these superlattices interesting in valley-electronics\cite{valley1,valley2,valley3}. These properties strongly suggest that the proposed TMDC multilayer-based superlattices are highly promising for optoelectronic and photovoltaic systems in the infrared range, which may promote an new research field for TMDC. Indeed, experimentally a similar superlattice material comprised of PbSe and MoSe$_2$ multilayers was successfully synthesized\cite{jacs-mo-pbse} and SOC induced VB splitting was experimental observed\cite{SOC-vbs}.

\begin{acknowledgments}
This work was financially supported by the National Natural Science Foundation of China (NSFC), Grant Nos. 11104191,11047104, 11004244 and 11274380, the Beijing Natural Science Foundation (BNSF), Grant No. 2112019, and the Basic Research Funds in Renmin University of China from the Central Government (Grant No. 12XNLJ03). W.J. was supported by the Program for New Century Excellent Talents in University. Y.Z. was supported by the Hong Kong University Grant Council (Contract No. AoE/P-04/08) of the Government of HKSAR. Calculations were performed at the Physics Lab of High-Performance Computing of Renmin University of China and Shanghai Supercomputer Center.
\end{acknowledgments}

\end{document}